 \documentclass [12pt,a4paper      ]{article}
\usepackage{graphics}
\usepackage{times}

\DeclareFontFamily{OT1}{times}{}
\DeclareFontShape {OT1}{times}{m }{n }{ <-> ptmr }{}
\DeclareFontShape {OT1}{times}{bx}{n }{ <-> ptmb }{}
\DeclareFontShape {OT1}{times}{m }{it}{ <-> ptmri}{}
\DeclareFontShape {OT1}{times}{bx}{it}{ <-> ptmbi}{}
\usepackage{amsmath}
\usepackage{amsfonts}
\usepackage{amssymb}
\usepackage{latexsym}
\newcommand{\cl}{C \kern -0.1em \ell} 
\newcommand{\bbH}{\mathbb{H}}         
\newcommand{\bbB}{\mathbb{B}}         
\newcommand{\bbO}{\mathbb{O}}         

\newcommand{\VEC}{\vec{\kern +.1em[}} 
\newcommand{\TOR}{\vec{\kern +.2em]}} 
\newcommand{\BRA}{\langle\kern -.2em\langle} 
\newcommand{\KET}{\rangle\kern -.2em\rangle} 

\newcommand{\Oh}{\tfrac{1}{2}}        
\newcommand{\Th}{\tfrac{3}{2}}        

\begin{document}

\title{\bf\vspace{-2.5cm} Quaternions in mathematical physics (2):\\
                             \emph{Analytical bibliography} }

\author{Andre Gsponer and Jean-Pierre Hurni\\ ~\\ 
\emph{Independent Scientific Research Institute}\\ 
\emph{Oxford, OX4 4YS, England}}

\renewcommand{\today}{6 July 2008}
\date{ISRI-05-05.26 ~~ \today}

\maketitle

\begin{abstract}

This is part two of a series of four methodological papers on (bi)quaternions and their use in theoretical and mathematical physics: 1~- Alphabetical bibliography, 2~- Analytical bibliography, 3~- Notations and definitions, and 4~- Formulas and methods.  

This quaternion bibliography will be further updated and corrected if necessary by the authors, who welcome any comment and reference that is not contained within the list.

\end{abstract}

~\\

\begin{center}
\resizebox{3cm}{!}{ \includegraphics{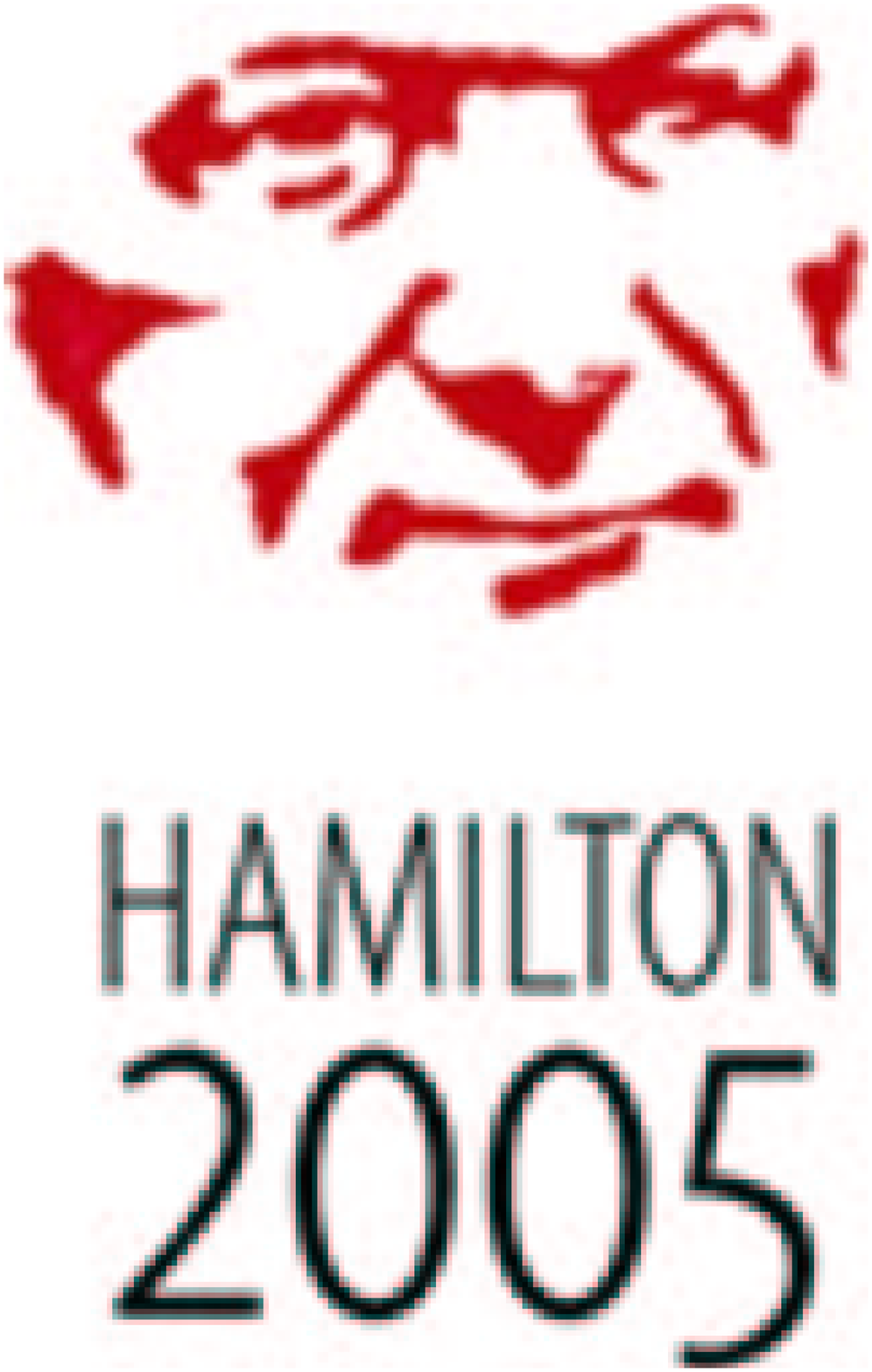}}
\end{center}

~\\

\noindent Living report, to be updated and corrected by the authors, first published on the occasion of the bicentenary of the birth of William Rowan Hamilton (1805--2005).
\newpage


\section{\Huge Table of contents}
\label{Table of contents}

The table of content is essentially the list of the main KEYWORDs (always written in the singular) used for classifying the items in the bibliography.

\renewcommand{\labelitemi}{}   
\renewcommand{\labelitemii}{}
\begin{itemize}

\item \ref{Table of contents}.         Table of contents 

\item \ref{Introduction}.              Introduction 

\item \ref{MATHEMATICS}.               MATHEMATICS
   \begin{itemize}
   \item \ref{MATH-VARIA}.             MATH-VARIA
   \item \ref{ALGEBRA}.                ALGEBRA
   \item \ref{INTEGRAL-QUATERNION}.    INTEGRAL-QUATERNION
   \item \ref{EQUATION}.               EQUATION
   \item \ref{LINEAR-FUNCTION}.        LINEAR-FUNCTION
   \item \ref{MATRIX}.                 MATRIX
   \item \ref{DETERMINANT}.            DETERMINANT
   \item \ref{GROUP-THEORY}.           GROUP-THEORY
   \item \ref{ANALYSIS}.               ANALYSIS
   \item \ref{ANALYTICITY-VARIA}.      ANALYTICITY-VARIA
   \item \ref{ANALYTICITY-H}.          ANALYTICITY-H
   \item \ref{ANALYTICITY-B}.          ANALYTICITY-B
   \item \ref{ANALYTICITY-CLIFFORD-R}. ANALYTICITY-CLIFFORD-R
   \item \ref{ANALYTICITY-CLIFFORD-C}. ANALYTICITY-CLIFFORD-C
   \item \ref{MANIFOLD}.               MANIFOLD
   \end{itemize}

\item \ref{RELATIVITICS}.          RELATIVISTICS
   \begin{itemize}
   \item \ref{SPECIAL-RELATIVITY}. SPECIAL-RELATIVITY
   \item \ref{CONFORMALITY}.       CONFORMALITY
   \item \ref{TENSOR}.             TENSOR
   \item \ref{SPINOR}.             SPINOR
   \item \ref{TWISTOR}.            TWISTOR
   \item \ref{GENERAL-RELATIVITY}. GENERAL-RELATIVITY
   \end{itemize}

\item \ref{FIELDS}.                        FIELDS
   \begin{itemize}
   \item \ref{SPIN-1}.                     SPIN-1 (MAXWELL, PROCA)
   \item \ref{SPIN-1/2}.                  SPIN-1/2 (DIRAC, LANCZOS, PAULI, WEYL)
   \item \ref{SPIN-3/2}.                   SPIN-3/2
   \item \ref{ANALYTICITY-MAXWELL}.        ANALYTICITY-MAXWELL
   \item \ref{ANALYTICITY-DIRAC}.          ANALYTICITY-DIRAC
   \end{itemize}

\item \ref{PHYSICS}.                        PHYSICS
   \begin{itemize}
   \item \ref{PHYSICS-VARIA}.               PHYSICS-VARIA
   \item \ref{MECHANICS}.                   MECHANICS
   \item \ref{HYDRODYNAMICS}.               HYDRODYNAMICS
   \item \ref{ELECTRODYNAMICS}.             ELECTRODYNAMICS
   \item \ref{LEPTODYNAMICS}.               LEPTODYNAMICS
   \item \ref{HADRODYNAMICS}.               HADRODYNAMICS
   \item \ref{PARTICLE-PHYSICS}.            PARTICLE-PHYSICS
   \end{itemize}

\item \ref{QUANTICS}.                        QUANTICS
   \begin{itemize}
   \item \ref{QUANTUM-PHYSICS}.              QUANTUM-PHYSICS
   \item \ref{QUANTUM-ELECTRODYNAMICS}.      QUANTUM-ELECTRODYNAMICS
   \item \ref{QUATERNIONIC-QUANTUM-PHYSICS}. QUATERNIONIC-QUANTUM-PHYSICS
   \end{itemize}

\item \ref{ALLIED FORMALISMS}.   ALLIED FORMALISMS
   \begin{itemize}
   \item \ref{OCTONION}.         OCTONION
   \item \ref{GRASSMANN}.        GRASSMANN
   \item \ref{CLIFFORD}.         CLIFFORD
   \item \ref{EDDINGTON}.        EDDINGTON
   \item \ref{SEMIVECTOR}.       SEMIVECTOR
   \item \ref{HESTENES}.         HESTENES
   \end{itemize}

\item \ref{MISCELLANEA}.     MISCELLANEA
   \begin{itemize}
   \item \ref{HISTORY}.      HISTORY and APPRECIATION
   \item \ref{BIBLIOGRAPHY}. BIBLIOGRAPHY
   \end{itemize}

\item \ref{Conventions}.     Conventions used in the bibliography
 
\end{itemize}
\renewcommand{\labelitemi}{$\bullet$}    
\renewcommand{\labelitemii}{\bfseries --}

\newpage

\section{\Huge Introduction}
\label{Introduction}

\emph{The purpose of the present analytical bibliography is to present a selection, but as comprehensive as possible, of the use of biquaternions in theoretical and mathematical physics, with an emphasis on their applications to fundamental rather than applied topics.}\footnote{The only exceptions to this rule are papers or books of general interest, and papers included for completeness when they are important to understand other papers.}   

This bibliography is already available as an alphabetic list \cite{GSPON2005D}. But, in that form, it is not much more useful than the internet.
What is needed is a logically sorted bibliography  --- an analytical bibliography that can be used by physicists to solve a problem, and by mathematicians to see what is of interest to physicists.

As is well known to anybody who has tried to classify a large set of scientific papers, sorting a bibliography is a very difficult and time-consuming task: It took us over ten years to bring this bibliography into its present form.  

Of course, there is a certain amount of subjectivity and arbitrariness in designing any classification scheme.  For this reason there will be a few sentences of introduction at the beginning of every subset of papers, explaining what is being collected in every chapter and section of the bibliography.  

Similarly, in this general introduction, we would like to explain what we mean by \emph{``mathematical physics,''} the concept which appears in the titles of this series of methodological papers \cite{GSPON2005D, GSPON2005E, GSPON2005F, GSPON2005G}, and which is our main thread in compiling and sorting our alphabetical and analytical bibliographies.

According to Ludvig Faddeev, \emph{``the main goal of mathematical physics is the use of mathematical intuition for the derivation of really new results in fundamental physics''} \cite{FADDE2000-}.  In the present case, the mathematical tool is complex quaternion algebra and analysis, which is so well suited to physics in our four-dimensional world that there is no important fundamental result which cannot be elegantly formulated and concisely derived using biquaternions, and only few quaternionic generalizations of fundamental theories which do not correspond to physical reality.  This is, in a forceful way, the confirmation of the validity of ``Hamilton's conjecture,'' the intuition that motivated Hamilton's dedication to quaternions, and their applications to physics, for most of the second half of his life (see Ref.~\cite{GSPON1993B} in Sec.~\ref{PHYSICS-VARIA}).

  However, this bibliography is not restricted to just papers in which quaternions or biquaternions are used explicitly: it also covers papers in which a hypercomplex coordinate-free whole-symbol system allied to quaternions is used (e.g., Clifford-numbers, Pauli vector-matrices, Eddington-numbers, semivectors, two-component spinors, twistors, etc.), and papers in which a quaternion or biquaternion structure plays a central role.  The kind of papers that are not included are those in which a strictly conventional matrix-type formalism is used (e.g., the Pauli- or Dirac-matrix formalisms), and papers which have not been published (or would not qualify to appear) in peer-reviewed journals.

  An important criterion used in compiling our bibliography is that it includes only papers which we have read, so that we were able to attach a few keywords to each entry in the reference list.  These keywords have the format \%\%KEYWORD, where ``\%'' is the symbol used for comments in \TeX~so that they  do not appear in the compiled bibliography, and where KEYWORD is always written in capital letters and in the singular.  However, the keywords are visible and can be searched for in the  \TeX-source of the bibliography.  As a matter of fact, this is how the present ``analytical'' bibliography was created starting from the ``alphabetical'' one.   For this reason the titles of the following chapters, Chaps.~\ref{MATHEMATICS} to \ref{MISCELLANEA}, and their sections, are nothing but the main keywords attached to every reference listed in them.

Finally, in Chap.~\ref{Conventions}, at the end of the bibliography, we detail the conventions used for the labels and styles of all types of references, of which typical examples are given in Sec.~\ref{Format}.


\section{\Huge MATHEMATICS}
\label{MATHEMATICS}

This chapter contains a selection of mathematical quaternion-papers which are of direct interest to mathematical physics. It not does contain however numerous papers or books in which quaternions, biquaternions, or quaternion structures are primarily studied or used in the context of ``pure mathematics.''

While the concept of quaternion as defined by Hamilton has a universal acceptance, it should be stressed that there are several definitions for related concepts such as ``biquaternions'' and ``pseudo-'' or ``generalized-'' quaternions.  As is explained in our paper on notations and terminology, Ref.~\cite{GSPON2005F} of Chap.~\ref{Introduction}, we remain as much possible consistent with Hamilton's original definitions.  For example, just like quaternions will always be elements $x \in \bbH$, the term biquaternion will always refer to Hamilton's complexified quaternions (i.e., $x\in \bbB$), and not to Cayley's (which are now called ``octonions,'' $x\in \bbO$), or to Clifford's (which are in a way an anticipation of Penrose's ``twistors'').\footnote{There are four Clifford algebras of dimension 8 over the reals: $\cl_{3,0}$ and $\cl_{1,2}$ are isomorphic to $\mathbb{B}$, whereas $\cl_{0,3}$ and $\cl_{2,1}$ are Clifford's misnamed ``biquaternions.''}

\subsection{MATH-VARIA}
\label{MATH-VARIA}

The books by W.R.\ Hamilton, P.G.\ Tait, and C.J.\ Joly listed in this section also contain chapters or sections on the applications of quaternions to classical physics.

\subsection{ALGEBRA}
\label{ALGEBRA}

Emphasis on algebraic operations and basic properties of algebras.  

See also GROUP-THEORY, Sec.~\ref{GROUP-THEORY}.

\subsection{INTEGRAL-QUATERNION}
\label{INTEGRAL-QUATERNION}

See also GROUP-THEORY, Sec.~\ref{GROUP-THEORY}.

\subsection{EQUATION}
\label{EQUATION}

See also DETERMINANT, Sec.~\ref{DETERMINANT}.

\subsection{LINEAR-FUNCTION}
\label{LINEAR-FUNCTION}

See also MATRIX, Sec.~\ref{MATRIX}.

\subsection{MATRIX}
\label{MATRIX}

See also DETERMINANT, Sec.~\ref{DETERMINANT}, EQUATION, Sec.~\ref{EQUATION}, and INTEGRAL-QUATERNION, Sec.~\ref{INTEGRAL-QUATERNION}.

\subsection{DETERMINANT}
\label{DETERMINANT}

See also MATRIX, Sec.~\ref{MATRIX}.

\subsection{GROUP-THEORY}
\label{GROUP-THEORY}

See also ALGEBRA, Sec.~\ref{ALGEBRA}.

\subsection{ANALYSIS}
\label{ANALYSIS}

Analysis, differentiation, integration, differential equations, differential forms, etc.

\subsection{ANALYTICITY-VARIA}
\label{ANALYTICITY-VARIA}

Various extensions of complex analyticity to spaces of higher dimensions.

\subsection{ANALYTICITY-H}
\label{ANALYTICITY-H}

Real quaternions analyticity only --- $\mathbb{H}$ alone

\subsection{ANALYTICITY-B}
\label{ANALYTICITY-B}

Biquaternions analyticity only --- $\mathbb{B}$ alone.

See also ANALYTICITY-MAXWELL, Sec.~\ref{ANALYTICITY-MAXWELL}, and ANALYTICITY-DIRAC, Sec.~\ref{ANALYTICITY-DIRAC}.

\subsection{ANALYTICITY-CLIFFORD-R}
\label{ANALYTICITY-CLIFFORD-R}

Real Clifford analysis --- but not $\mathbb{H}$ alone

\subsection{ANALYTICITY-CLIFFORD-C}
\label{ANALYTICITY-CLIFFORD-C}

Complex Clifford analysis --- but not $\mathbb{B}$ alone

\subsection{MANIFOLD}
\label{MANIFOLD}

Quaternionic manifolds, K\"ahlerian manifolds, symmetric-spaces, etc.

\section{\Huge RELATIVISTICS}
\label{RELATIVITICS}

Mathematical aspects of relativistics, i.e., relativity theory, and mathematical methods associated with that theory.  Papers dealing with specific applications to fields and physics are collected in Chaps.~\ref{FIELDS} and~\ref{PHYSICS}.

The exception to this rule is Sec.~\ref{GENERAL-RELATIVITY} on general relativity theory which does not allow for such a separation since according to it gravitation is a purely geometrical effect.

\subsection{SPECIAL-RELATIVITY}
\label{SPECIAL-RELATIVITY}

Lorentz transformations and representations of the Lorentz group (rotations, boosts, and improper transformations), as well as more general rotations in Minkowskian or Euclidian four-space, including discrete transformations.

See also ELECTRODYNAMICS, Sec.~\ref{ELECTRODYNAMICS}, in particular for the seminal papers of A.W.\ Conway and L.\ Silberstein.

\subsection{CONFORMALITY}
\label{CONFORMALITY}

Mostly mathematical papers dealing with conformal-transformations and conformal-invariance in hyperbolic four-space.  

See also TWISTOR, Sec.~\ref{TWISTOR}.

\subsection{TENSOR}
\label{TENSOR}

Relations between quaternions and tensors.

\subsection{SPINOR}
\label{SPINOR}

Relations between quaternions and spinors.

\subsection{TWISTOR}
\label{TWISTOR}

Twistors may be regarded as spinors of the $O(4,2)$ group, which is two-to-one isomorphic with the full 15-parameter conformal group in Minkowski space, including the full Poincar\'e group.  As 8-dimensional points in complexified space-time, twistors are able to encode angular-momentum/spin in addition to position/translation. They were introduced by Roger Penrose as possibly more fundamental physical objects than 4-dimensional points in Minkowski space-time.

\subsection{GENERAL-RELATIVITY}
\label{GENERAL-RELATIVITY}

Papers on general relativity, and papers in which fields (e.g., Dirac, Maxwell) are expressed in curved space-time.


\section{\Huge FIELDS}
\label{FIELDS}

This chapter contains papers related to classical (i.e., ``non-quantized'') fields which have a stronger emphasis on mathematics and theory than physics and applications. 

Papers in which fields are expressed in curved space-time are collected in Sec.~\ref{GENERAL-RELATIVITY}.

Papers dealing with specific applications are collected in Chap.~\ref{PHYSICS} or~\ref{QUANTICS}.  For example, Sec.~\ref{SPIN-1/2} in the present chapter contains theoretical papers on Dirac's equation and field, while Sec.~\ref{QUANTUM-PHYSICS} contains papers in which Dirac's equation is applied to atomic physics.

The use of biquaternions naturally implies formulations in which Minkowski's metric and Einstein's relativity are automatically implemented.  However, papers in which Galilean relativity and non-relativistic limits are considered can also be written using quaternions or biquaternions in such a way that the relations to Minkowski's metric and Einstein's relativity are hidden or lost.  For example, Pauli's equation for a non-relativistic spin~$\Oh$ field can be written as a purely real quaternion equation.  Such papers are also collected in this chapter.

\subsection{SPIN-1 (MAXWELL, PROCA)}
\label{SPIN-1}

This section contains papers on Maxwell's and Proca's equations and fields, as well as papers on spin~0 fields when they are discussed in conjunction with spin~1 fields.

Applications are in sections ELECTRODYNAMICS and QUANTUM-ELEC\-TRODYNAMICS, i.e., Secs.~\ref{ELECTRODYNAMICS} and~\ref{QUANTUM-ELECTRODYNAMICS}.

See also ELECTRODYNAMICS, Sec.~\ref{ELECTRODYNAMICS}, for the seminal papers of A.W.\ Conway and L.\ Silberstein.

\subsection{SPIN-1/2 (DIRAC, LANCZOS, PAULI, WEYL)}
\label{SPIN-1/2}

Dirac's, Lanczos's, and Weyl's relativistic equations and fields, as well as non-relativistic equations and fields such as Pauli's.  

Papers such as Lanczos's which relates to fields with spin~0 to~$\Th$ but whose main emphasis is spin $\Oh$ are included in this section.

Applications are in sections QUANTUM-PHYSICS and QUANTUM-EL\-EC\-TRODYNAMICS, i.e., Secs.~\ref{QUANTUM-PHYSICS} and~\ref{QUANTUM-ELECTRODYNAMICS}. \ref{QUANTUM-ELECTRODYNAMICS}.

See also EDDINGTON and SEMIVECTOR, Secs.~\ref{EDDINGTON} and~\ref{SEMIVECTOR}.

\subsection{SPIN-3/2}
\label{SPIN-3/2}

\subsection{ANALYTICITY-MAXWELL}
\label{ANALYTICITY-MAXWELL}

Hypercomplex analysis applied to Maxwell's field, with emphasis on physics.

\subsection{ANALYTICITY-DIRAC}
\label{ANALYTICITY-DIRAC}

Hypercomplex analysis applied to Dirac's field.

\section{\Huge PHYSICS}
\label{PHYSICS}

This chapter contains papers in which the quaternion formalism and quaternion methods are applied to physics, with an emphasis on applications of a fundamental rather than practical character.

However, papers in which quaternions are applied to GENERAL-RELATIVITY are in Chap.~\ref{RELATIVITICS} on ``relativistics,'' and papers in which quaternions are applied to QUANTUM-PHYSICS are in Chap.~\ref{QUANTICS} on ``quantics.''

Moreover, as was explained in Chap.~\ref{Introduction}, the bibliography also includes selected papers in which a formalism allied to quaternions is used (e.g., semivectors or Clifford numbers), especially if these papers could have been written using quaternions rather than the closely related formalism. 

On the other hand, the numerous papers which use a standard tensor or matrix formalism (e.g., the Pauli- or Dirac-matrices, and the corresponding two- or four-component formalisms) are excluded from the bibliography.  The reason for this exclusion is conceptual rather than conventional (because Pauli's matrices are actually very closely related to Hamilton's quaternion units).  The reason is that the emphasis of standard formalisms is on the vector-space structure supported by them, while the emphasis of the biquaternion formalism is on the full algebraic structure (vector-space and multiplication ring) provided by the biquaternion algebra.

The instances in which selected papers using Pauli-matrices and two-component spinors are included in the bibliography (e.g., by F.\ G\"ursey, L.M.\ Brown, and even by Pauli himself) correspond to cases in which their use is essentially equivalent to that of biquaternions due to the isomorphism $\mathbb{B} \equiv  
M_2(\mathbb{C})$.

\subsection{PHYSICS-VARIA}
\label{PHYSICS-VARIA}

Papers and books dealing with several applications to physics, and few selected references dealing with applications in computer graphics, modeling, etc.

See also the books by W.R.\ Hamilton, P.G.\ Tait, and C.J.\ Joly listed in the MATH-VARIA section, Sec.~\ref{MATH-VARIA}, which contain chapters or sections on the applications of quaternions to classical physics topics such as mechanics, hydrodynamics, astronomy, etc..

\subsection{MECHANICS}
\label{MECHANICS}

This section also includes papers on ``analytical mechanics,'' and ``generalized dynamics,'' etc.

\subsection{HYDRODYNAMICS}
\label{HYDRODYNAMICS}

\subsection{ELECTRODYNAMICS}
\label{ELECTRODYNAMICS}

Classical electrodynamics.

\subsection{LEPTODYNAMICS}
\label{LEPTODYNAMICS}

Weak interactions of leptons and quarks, electroweak models and interactions.

\subsection{HADRODYNAMICS}
\label{HADRODYNAMICS}

Strong interactions, piondynamics, quantum-chromodynamics (QCD).

\subsection{PARTICLE-PHYSICS}
\label{PARTICLE-PHYSICS}
 
Papers dealing with elementary-particles physics and field-theory, as well as mathematical-physics papers primarily motivated by their relevance to elementary particle physics. 

Papers using OCTONION structures and fields are collected in Sec.~\ref{OCTONION}.

\section{\Huge QUANTICS}
\label{QUANTICS}

Quantum theory and its generalizations.

\subsection{QUANTUM-PHYSICS}
\label{QUANTUM-PHYSICS}

Standard relativistic and non-relativistic quantum physics in which the scalars belong to the field of complex numbers.

\subsection{QUANTUM-ELECTRODYNAMICS}
\label{QUANTUM-ELECTRODYNAMICS}

Quantum relativistic perturbation-theory of electromagnetic interactions.

Most papers relate to the attempt of R.P. Feynman's student L.M. Brown to reformulate quantum electrodynamics using, instead of the usual four-component-spinor Dirac matrix formalism, the two-component-spinor Pauli-matrix formalism which is essentially equivalent to that of biquaternions due to the isomorphism $\mathbb{B} \sim M_2(\mathbb{C})$.

\subsection{QUATERNIONIC-QUANTUM-PHYSICS}
\label{QUATERNIONIC-QUANTUM-PHYSICS}

Generalized quantum physics in which real quaternions are used instead of complex numbers as the scalar field. 

In the \TeX source of the bibliography the keyword QUATERNIONIC-QUANTUM-PHYSICS is abreviated QQPH, and there two special keywords: QQPH/CERN for the publications of the CERN-University-of-Geneva school (led by J.M.\ Jauch) and QQPH/ALDER for the papers of S.L.\ Adler and his followers.

\section{\Huge ALLIED FORMALISMS}
\label{ALLIED FORMALISMS}

This chapter contains papers on algebraic formalisms that are closely related to quaternions and biquaternions, either because they are essentially equivalent to them (e.g., Einstein-Mayer's semivectors) or a generalization of them (e.g., Clifford numbers).

However, this chapter covers only the major formalisms allied to quaternions, i.e., only those which have gained some level of popularity, and not the many formalisms that have been, and continue to be, introduced.

Since these formalisms have been invented after the discovery of quaternions, the related sections are listed in the historical order in which they were introduced.\footnote{For an historical account, see Ref.~\cite{VANDE1985-} in Sec.~\ref{HISTORY}.}

\subsection{OCTONION (``Cayley's numbers'')}
\label{OCTONION}

Octonions were independently discovered, first in 1843 a few months after the discovery of quaternions by Charles Grave (who called them ``octaves''), and then in 1845 by Arthur Cayley (who called them ``biquaternions'').

\subsection{GRASSMANN (``Grassmann's calculus of extensions'')}
\label{GRASSMANN}

Hermann G. Grassmann's ``calculus of extensions'' of 1844/1862 is the first attempt to define explicitly the notions of ``$n$-dimensional vector space,'' and of ``internal'' and ``external'' products of vectors.

\subsection{CLIFFORD (``Clifford numbers'')}
\label{CLIFFORD}

William K. Clifford introduced in 1873 two different kind of 8-dimensional algebras that he called ``biquaternions,'' and then in 1878 a general theory of linear associative algebras of dimension $2^n$, denoted $\cl_{p,q}$ where $n=p+q$.

\subsection{EDDINGTON (``Eddington numbers'')}
\label{EDDINGTON}

 The ``E-numbers'' introduced by Arthur S. Eddington in 1928 are equivalent to elements of the algebra of the Dirac matrices, i.e., to elements of the Clifford algebra $\cl_{4,1}$.  Eddington's formalism corresponds to the first use of Clifford numbers in relation to Dirac's theory.

\subsection{SEMIVECTOR (``Einstein-Mayer's spinors'')}
\label{SEMIVECTOR}

Semivectors were introduced in 1932 by Einstein and his assistant Walter Mayer, possibly in reaction to the work on Dirac's equation by his previous assistant, Cornelius Lanczos.  Einstein-Mayer's formulation of Dirac's equation is based on $4 \times 4$ matrices which in flat spacetime reduces to Lanczos's biquaternionic formulation.

\subsection{HESTENES (``Hestenes's space-time algebra'')}
\label{HESTENES}

Hestenes's 16-dimensional ``space-time algebra'' formalism (based on the Clifford algebra $\cl_{1,3}$) was designed in 1966 to provide a substitute for the standard 32-dimensional Dirac formalism (equivalent to the Clifford algebra $\cl_{4,1}$) such that all references to complex numbers are avoided.

\section{\Huge MISCELLANEA}
\label{MISCELLANEA}

\subsection{HISTORY and APPRECIATION}
\label{HISTORY}

Historical papers on quaternions and on Hamilton (1805-1865), as well as some appreciations of quaternions and Hamilton's devotion to them.

Historical papers on algebras in general. 

The historical papers on specific formalisms allied to quaternions are in the respective sections of Chap.~\ref{ALLIED FORMALISMS}.

\subsection{BIBLIOGRAPHY}
\label{BIBLIOGRAPHY}

\newpage

\section{\Huge Conventions used in the bibliography}
\label{Conventions}

\subsection{Conventions for the \TeX ~citation-label}
\label{Label}

The \TeX ~{\bf citation-label} has always 10 characters, i.e.:

~~~ ~~~ {\bf LASTN1234-} ~~~ or ~~~ {\bf LASTN1234X} ~,

\noindent were LASTN are the 5 first characters of the last name, ``1234'' the four-digit year, and ``X'' is A, B, C, ..., Z if there are more than one reference for the author in that year. 

If the author last name has less that 5 characters, the missing characters are replaced by ``-''s.

{\bf N.B.:} In the case where different authors have the same last name the 10th character (i.e., ``-'') will be used to remove possible ambiguities that may arise in a given year.

\subsection{Conventions for the references's "style"}
\label{Style}

~~~~~~Authors: Only first character of first-names. Ex: P.A.M.\ Dirac.

Composite last-names are concatenated: De Broglie  $\rightarrow$ DeBroglie.

Two authors: Use ``and'' between their names.

Three or more authors: Use ``, and'' before the last author in the list.

Book titles:    Capitalize First Letter of Each Word. Normal characters.

Article titles: Capitalize only first letter of first word. \emph{Italic} characters.

Editor(s): Use  ``ed.'' or  ``eds.''

Latin: Use only unambiguous combinations such as:  ``et al.'', ``ibid.'', `in:''...

Volume numbers: {\bf Bold} characters.

No other bold symbol should appear anywhere in the bibliography file.

Number of pages in a book:  ``123~pp.'' (Non-break space: ``$\sim$'' in~ \TeX.)

First and last pages of article: ``123--456" (Medium dash: ``-\,-'' in~ \TeX.)

Examples of typical references are given in Sec.~\ref{Format}.

\subsection{Internet URLs and {\tt arXiv.org} links}
\label{internet}

In order that they are properly processed by the {\tt arXiv.org} \TeX-compiler, a white space ``\underline{ }'' is put in the front and in the back of all internet URLs and {\tt arXiv.org} links.  Moreover, they are \underline{ underlined }.  These conventions insure that the URLs and links are properly recognized, and that they are not hyphenated.  

For instance, clicking the {\tt arXiv.org} link \underline{ arXiv:name/number } automatically generates the internet URL \underline{ http://www.arXiv.org/abs/name/number }.

\subsection{Non-\TeX ~conventions: \emph{directives} and \emph{keywords}}
\label{Non-TeX}

In the file special \TeX ~comments are used to introduce the keywords, and to introduce directives to the bibliography update/search/display computer programs:

\begin{itemize}
\item Standard comments start with ``\%''
\item Bibliography KEYWORDS   start with ``\%\%''
\item Bibliography directives start with ``\%\$''
\end{itemize}

\subsection{Directives}
\label{Directives}

\begin{itemize}
\item \%\$D : date of entry in DDMMYYYY format
\item \%\$C : a commentary/appreciation of the document follows
\item \%\$L : this reference is on loan to the person whose name follows
\item \%\$M : this reference is missing or lost
\item \%\$N : this reference is not yet filed in its envelope/folder
\item \%\$O : this reference is on order from some library
\item etc. 
\end{itemize}

The date of entry is particularly important for making automatic updates.  It is normally the last item in a bibliography entry, unless there is a commentary which is appended at the very end of it.

\subsection{Format of typical references}
\label{Format}

What follows is a list of typical references in which the label, keywords, and directives are made visible.

\begin{itemize}

\item {\bf Book:}

{\bf LANCZ1949-} C. Lanczos, The Variational Principles of Mechanics (Dover, New-York, 1949, 1986) 418~pp.  Quaternions pages 303--314. \%\%BOOK, \%\%QUATERNION, \%\%SPECIAL-RELATIVITY, \%\$D06022002.

\item {\bf Conference proceedings, festschrift or contributed volume:}

{\bf SPROS1996-} W. Spr\"ossig and K. G\"urlebeck, eds., Proc. of the Symp. ``Analytical and Numerical Methods in Quaternionic and Clifford Analysis,'' Seiffen, June 5--7, 1996 (TU Bergakademie Freiberg, 1996) 228~pp.  \%\%BOOK, \%\%QUATERNION, \%\%ANALYTICITY, \%\%CLIFFORD, \%\$D20032002.

\item {\bf Paper in a conference proceedings, festschrift, or contributed volume:}

{\bf PENRO1990-} R. Penrose, \emph{Twistors, particles, strings and links}, in: D.G. Quillen et al., eds., The Interface of Mathematics and Particle Physics (Clarendon Press, Oxford, 1990) 49--58. \%\%QUATERNION, \%\%TWISTOR, \%\$D05022002.

\item {\bf Paper in a journal:}

{\bf WEISS1941-} P. Weiss, \emph{On some applications of quaternions to restricted relativity and classical radiation theory}, Proc. Roy. Irish Acad. {\bf A 46} (1941) 129--168. \%\%QUATERNION, \%\%MAXWELL, \%\%SPINOR, \%\%LORENTZ-DIRAC, \%\$D09022002.

\item {\bf Two authors:}

{\bf EINST1932-} A. Einstein and W. Mayer, \emph{Semi-Vektoren und Spinoren}, Sitzber. Preuss. Akad. Wiss. Physik.-Math. Kl. (1932) 522--550. \%\%QUATERNION, \%\%SEMIVECTOR, \%\$D05022002.

\item {\bf Report or preprint:}

{\bf VELTM1997-} M. Veltman, \emph{Reflexions on the Higgs system}, Report 97-05 (CERN, 1997) 63~pp. \%\%BOOK, \%\%QUATERNION, \%\%LEPTODYNAMICS, \%\%GAUGE-THEORY,  \%\$D14052001, \%\$C Veltman uses quaternions in the form of 2x2 matrices.

\item {\bf Ph.D. thesis:}

{\bf GURSE1950A} F. G\"ursey, \emph{Applications of quaternions to field equations}, Ph.D.\ thesis (University of London, 1950) 204~pp.  \%\%BOOK, \%\%QUATERNION, ~~\%\%DIRAC, ~~\%\%GENERAL-RELATIVITY, ~~\%\%LANCZOS, \%\%PROCA, \%\$D20032002. 

\item {\bf Unpublished document}:

{\bf GSPON1993A} A. Gsponer and J.-P. Hurni,  \emph{Quaternion bibliography: 1843--1993}. Report ISRI-93-13 (17 June 1993) 35~pp. This is the first version of the present bibliography, with 228 entries. \%\%QUATERNION, \%\%BIBLIOGRAPHY, \%\$D20032002.

\item {\bf ``arXived'' e-print:}

{\bf GSPON2002D} A. Gsponer, \emph{Explicit closed-form parametrization of $SU(3)$ and $SU(4)$ in terms of complex quaternions and elementary functions}, Report ISRI-02-05 (22 November 2002) 17~pp.;  e-print \underline{ arXiv:math-ph/0211056 }.  \%\%QUATERNION, \%\%ALGEBRA, \%\%HADRODYNAMICS, \%\$10092005.

\item {\bf Internet page:}

{\bf NOBIL2006-} R. Nobili, \emph{Fourteen steps into quantum mechanics}, HTML document (Posted in 2006) about 13~pp.;  available at\\ \underline{ http://www.pd.infn.it/~rnobili/qm/14steps/14steps.htm }. \%\%QUATERNION, \%\%QUANTUM-PHYSICS, \%\%QQM, \%\$D25.

\end{itemize}

\section*{ACKNOWLEDGMENTS}
\label{ACKNOWLEDGMENTS}

This bibliography would not exist without the help, dedication, and professionalism of Mrs.\ Claire-Lise Held and Mrs.\ Jocelyne Favre at the University of Geneva Physics department library, and many other librarians at other universities.

\end{document}